\documentclass[journal]{IEEEtran}
\usepackage{xcolor}
\usepackage{balance}
\usepackage[pdftex]{graphicx}

\usepackage{amsmath}
\usepackage{amsthm,amsfonts}
\usepackage[greek,english]{babel}
\usepackage{circuitikz}
\usepackage{pgfplots}
\usepackage{url}

\newcommand{\gpi}{\textrm{\greektext p}}
\renewcommand{\pi}{\gpi}
\providecommand*{\mat}[1]{\mathbf#1}
\providecommand*{\M}[1]{\mathbf#1}
\providecommand*{\mrm}[1]{\mathrm{#1}}
\providecommand*{\T}[1]{\mathrm{#1}}

\DeclareMathAccent{\ring}{\mathalpha}{operators}{"17}

\providecommand*{\ju}{\ensuremath{\mrm{j}}}

\newcommand{\eig}{\mathop{\mrm{eig}}}

\newcommand{\reg}{\varOmega}

\newcommand{\herm}{\mrm{H}}
\newcommand{\atan}{\mathop{\mrm{atan}}}

\newcommand{\ie}{\textit{i.e.}\/, }
\newcommand{\eg}{\textit{e.g.}\/, }

\newcommand{\minimize}{\mrm{minimize}}

\newcommand{\maximize}{\mrm{maximize}}

\newcommand{\subto}{\mrm{subject\ to}}

\colorlet{dpurple}{blue!50!red}
\colorlet{dblue}{blue!50!black}
\colorlet{dgreen}{green!50!black}
\colorlet{dred}{red!50!black}
\colorlet{dyellow}{yellow!50!black}
\colorlet{dorange}{orange!50!black}
\definecolor{metal}{RGB}{218,165,32}
\definecolor{diel}{RGB}{1,165,32}
\definecolor{antenna}{RGB}{100,150,162}
\definecolor{reg}{rgb}{0.2,0.6,0.8}%
\definecolor{preg}{rgb}{0.8,0.2,0.2}%

\tikzset{>=latex}

\begin{document}

\pagestyle{headings}

\title{Fundamental Limits of Characteristic Mode Slopes}

\author{Johan Lundgren, \IEEEmembership{Member, IEEE}
 and Mats Gustafsson, \IEEEmembership{Senior Member, IEEE}
\thanks{Manuscript received \today; revised \today. This work was supported the Swedish Research Council SEE-6GIA and SSF Sabbatical.}
\thanks{J. Lundgren and M. Gustafsson are with Electrical and Information Technology, Lund University, Lund, Sweden, (e-mails: \{johan.lundgren,mats.gustafsson\}@eit.lth.se).}
}

\maketitle

\begin{abstract}
Characteristic Mode analysis is a widely used technique in antenna design, providing insight into the fundamental electromagnetic properties of radiating structures. In this paper, we establish fundamental bounds on the slope of characteristic mode eigenvalues and angles, demonstrating that their rate of change is subject to fundamental constraints for all possible realizations within a given design region. These bounds are derived using the method of moments and reformulating the frequency derivative (slope) of the eigenvalue quantities as an optimization problem over the current distribution confined to the design region. The results reveal a direct analogy between these constraints and classical antenna Q-factors, highlighting the intrinsic limitations on modal evolution and their implications for bandwidth and miniaturization in antenna design. Moreover, by iteratively enforcing orthogonality among the modes the derived bounds can be tightened for higher-order modes, providing deeper insight into the number of simultaneous, usable modes and their associated degrees of freedom. These bounds provide a feasibility criterion for achievable modal behavior, offering insights that can guide the design process. Examples are given for various surface PEC structures. 
\end{abstract}

\vskip0.5\baselineskip

\section{Introduction}
Antennas have been a cornerstone of modern society for over a century, evolving alongside both technological innovations, shifting design requirements, and our deepening understanding of electromagnetic theory. In the pioneering era of wireless communication, following Hertz and Marconi, antennas were necessarily large structures operating at low frequencies, thus electrically small, due to both technological limitations and the fundamental physics~\cite{Balanis2005,Volakis+etal2010}. As technology advanced, higher frequencies became accessible, enabling more compact and integrated antenna solutions~\cite{Volakis+etal2010}.

With these new possibilities, designers became increasingly interested in understanding the fundamental limits of antenna performance - what could ultimately be achieved given specific constraints. While some design challenges can be addressed through innovation, others are dictated by fundamental physical principles. For the past 75 years, extensive research has focused on defining these constraints. Early works by Chu~\cite{Chu1948} and Wheeler~\cite{Wheeler1947} established the initial bounds for antennas in terms of bandwidth limitations and Q-factors. Subsequent work continued to establish rigorous bounds on antenna performance, providing insight into the trade-offs and limitations for antennas~\cite{Hansen1981,Yaghjian+Best2005,Gustafsson+etal2007a,Gustafsson+etal2012a}.

One well-known challenge in antenna design is the inherently narrow bandwidth of electrically small antennas, a limitation dictated by fundamental physics~\cite{Volakis+etal2010}. Further research has led to tight, geometry- and material-dependent bounds for many important antenna configurations, defining the ultimate limits of what is physically achievable~\cite{Gustafsson+etal2007a,Gustafsson+etal2015b,Capek+etal2019b}. 

However, translating these theoretical limits into practical design insights is not always straightforward, as real-world performance and metrics is also influenced by ongoing technological advancements and evolving design methodologies.

A widely adopted approach in antenna design and analysis is Characteristic Mode Analysis (CMA), which provides a systematic framework for revealing the fundamental electromagnetic behavior of conducting and dielectric structures. Originally introduced by Garbacz~\cite{Garbacz1965} and later popularized by Harrington and Mautz~\cite{Harrington+Mautz1971}, CMA provides a set of modes intrinsic to the geometry and material composition of the object under study. These modes are typically computed using the Method of Moments (MoM), though alternative numerical techniques based on electromagnetic scattering formulations are also applicable~\cite{Garbacz+Turpin1971,Gustafsson+etal2022a,Capek+etal2023a}. The resulting modal decomposition, typically expressed in terms of surface currents and, in the case of lossless systems, orthogonal far fields, reveals how a structure naturally supports radiation at a given frequency. This provides physical insight that is valuable for antenna synthesis, optimization, and performance evaluation~\cite{Chen+Wang2015,Cabedo-Fabres+etal2007,Lau+etal2022}.

By examining characteristic mode quantities such as eigenvalue traces $\lambda_n(\omega)$, characteristic angles $\alpha_n(\omega)$, and far-field patterns across a frequency range, CMA enables detailed tracking of modal behavior and evolution~\cite{Chen+Wang2015,Cabedo-Fabres+etal2007}. A key advantage of the method is its ability to isolate dominant radiating modes, supporting targeted geometry tuning for optimal excitation and impedance matching~\cite{Luomaniemi+etal2021,Martens+Manteuffel2014,Li+etal2022}.

Since characteristic modes reflect the underlying physics that govern antenna radiation, they must adhere to the same fundamental performance bounds as the antenna itself. This implies that modal quantities are linked to other well-known figures of merit in antenna theory. For instance, the relationship between CM eigenvalues and Q-factors provides a direct link between modal properties and bandwidth limitations, highlighting the relevance of CMA in evaluating and optimizing practical antenna designs~\cite{Chalas+etal2016,Capek+Jelinek2016}. However, while such relationships are understood, specific bounds on the characteristic modes themselves have yet to be presented. In this paper, we establish bounds on the slope (frequency derivative) of the characteristic eigenvalues and angles. These bounds are not limited to any specific realization but apply universally for all structures within a defined design region, encompassing all possible characteristic values, regardless of whether a known realization exists or not.

To facilitate this discussion, Sec.~\ref{sec:CMintro} provides a brief introduction to CM analysis within the MoM framework, followed by a motivation for establishing bounds at resonance in Sec.~\ref{sec:resonance}. The bounds are then established for the general case in Sec.~\ref{sec:general}, with results for various geometries in Sec~\ref{sec:results} and extension to higher order modes in Sec.~\ref{Sec:HOM}. The work is concluded in Sec.~\ref{sec:conc}.

\section{Characteristic mode overview}
\label{sec:CMintro}
Every object interacts with the electromagnetic field based on factors such as geometry and material properties. A key interest is to understand and characterize conditions where the object exhibits a strong interaction with the field—whether through scattering (incoming fields) or radiation (outgoing fields). These strong interactions are intrinsic to the object's structure and are precisely what characteristic modes aim to describe. They can be computed using impedance-based formulations (e.g., MoM) or extracted from scattering matrices via numerical methods such as Finite Element Method (FEM) and Finite-Difference Time-Domain (FDTD)~\cite{Garbacz+Turpin1971,Gustafsson+etal2022a,Capek+etal2023a}. The implementation and numerical difficulty varies and there are drawbacks  and gains with each formulation. In this work we utilize MoM and study lossless materials. The object is then described by the matrix relation 
\begin{equation}
    \mat Z \mat I=\mat V,
\end{equation}
where $\mat Z$ represents the electric field integral equation MoM matrix~\cite{Harrington1968} for a region $\varOmega$. The characteristic modes are obtained through the solution to the generalized eigenvalue problem~\cite{Harrington+Mautz1971},
\begin{equation}
    \mat X \mat I_n=\lambda_n \mat R\mat I_n, \quad \text{where}\quad \mat Z=\mat R+\ju\mat X.
    \label{eq:CM}
\end{equation}
The generalized eigenvalue problem formulation can be interpreted as a decomposition of the structure's response into reactive ($\mat X$) and resistive ($\mat R$) components. 

In addition to the field distribution obtained in standard numerical solvers, CMA offers deeper insight into how a structure supports radiation and scattering~\cite{Chen+Wang2015,Cabedo-Fabres+etal2007,Lau+etal2022}.

\begin{figure}
    \centering
    \includegraphics[width=0.95\linewidth]{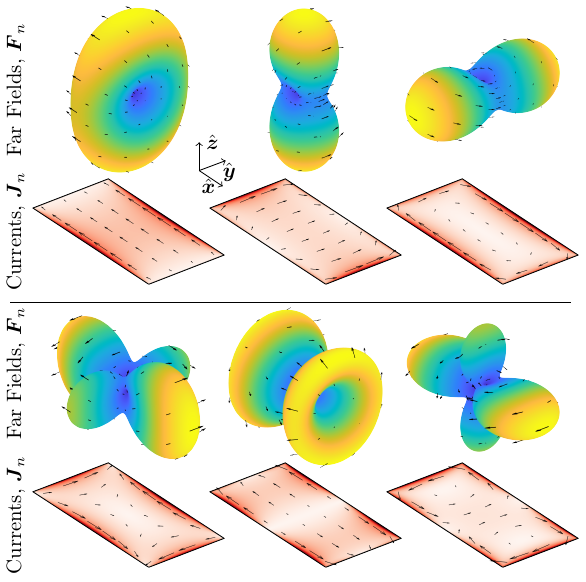}
    \caption{Characteristic currents at their respective resonances and the corresponding far fields for a PEC plate with a 2:1 aspect ratio}
    \label{fig:CharQuant}
\end{figure}

The characteristic modes can be used as a basis inherent to the structure where a total current can be expressed as a weighted sum of characteristic currents, each associated with a specific eigenvalue~\cite{Harrington+Mautz1971}. Several important quantities—such as characteristic far fields, currents, eigenvalues, angles, and modal significances—can be determined from the generalized eigenvalue equation~\eqref{eq:CM}. In Fig.~\ref{fig:CharQuant} a PEC plate with an aspect ratio of 2:1 is displayed, along with the characteristic currents for the first six modes at their respective resonance along with their associated characteristic far fields. We observe that these modes include familiar dipole-like currents and far fields, as well as loop modes and higher-order modes. These modes form the CM basis for this particular structure. They naturally occur on the structure and can be excited and combined by antenna designers to create efficient antenna designs. For other designs, the current distributions and far-field patterns may be more elaborate. 

Commonly, the key parameters used to describe the behavior and usefulness of the CMs are the characteristic eigenvalues $\lambda_n$, characteristic angles $\alpha_n$, and modal significance $|t_n|$, which are related through:
\begin{equation}
    \alpha_n=\mathrm{arg}(t_n)=\pi-\atan (\lambda_n), \quad |t_n|=\frac{1}{\sqrt{1+\lambda_n^2}}.
    \label{eq:Quantrel}
\end{equation}
In this work we focus on the characteristic angles, $\pi/2<\alpha_n<3\pi/2$, where a resonant mode is described by $\alpha_n=\pi$, inductive behavior occurs  when $\pi/2 < \alpha_n<\pi$ and capacitive behavior for $\pi <\alpha <3\pi/2$. 

Since modal quantities vary continuously with frequency, except in cases where degenerate modes occur, it is often useful for antenna designers to analyze their behavior to identify for instance broadband capabilities of modes, characterized by $\alpha \approx \pi$ over a wide frequency range. Degeneracies arise due to geometrical symmetries~\cite{Wigner+Neumann1929,Schab+Bernhard2016}, and can be treated by decomposing the CM formulation into orthogonal subspaces that are free from degenerate modes, as assumed here.

To capture the essential frequency dependence with respect to the physical size of the structure, this work adopts the dimensionless parameter $ka$, where $k=\omega/c$ denotes the free-space wavenumber, $c$ speed of light, and $a$ is the radius of the smallest sphere that fully encloses the design region $\varOmega$. This normalization facilitates a general characterization of the system behavior across different frequency regimes.

\begin{figure}
    \centering
    \includegraphics[width=0.95\linewidth]{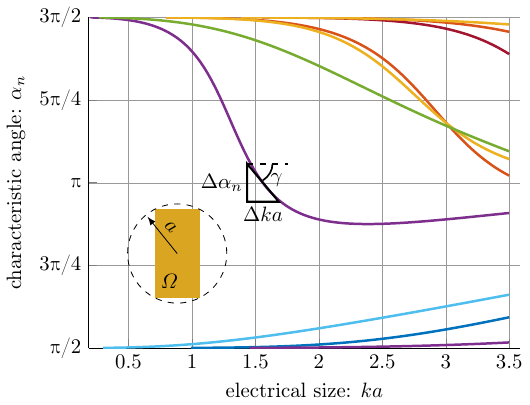}
    \caption{Characteristic angles of a 2:1 PEC plate realization $\reg_\T r$, using the full design space $\reg$. The design region is circumscribed by a sphere of radius $a$. The slope angle ($\gamma$) and the slope (black triangle) of the characteristic angle at first resonance is highlighted.}
    \label{fig:2t1plateModes}
\end{figure}

Returning to the 2:1 aspect ratio PEC plate in Fig.~\ref{fig:CharQuant}, let $\reg$ denote the region it occupies. Its characteristic modes are represented through the characteristic angles for various electrical sizes (or equivalently, frequencies) in Fig.~\ref{fig:2t1plateModes}. The region $\reg$ is circumscribed by a sphere of radius $a$. Several modes are displayed, showing the evolution of the characteristic angles as the electrical size of the object increases.  From elementary antenna theory, we expect this geometry to support a dipole-like mode when its longest side corresponds to half the operating wavelength, as seen in Fig.~\ref{fig:CharQuant}. This condition occurs when the electrical size is $ka = \sqrt{5}\pi/4\approx 1.76$, which coincides with the first mode reaching resonance ($\alpha = \pi$). This mode, purple line in Fig.~\ref{fig:2t1plateModes}, originates from the capacitive side ($\alpha=3\pi/2$). Conversely, initial inductive modes appear from $\alpha=\pi/2$, of which three can be seen in Fig.~\ref{fig:2t1plateModes}, all of which are far from resonance.

A mode near resonance, $\alpha_n\approx\pi$, corresponds to a well-functioning mode for antenna applications over that frequency range. For a wide-band application of a single mode the slope of $\alpha_n$ should then be as flat as possible around this range. In Fig.~\ref{fig:2t1plateModes}, the slope at resonance is indicated by the black triangle, showing a negative slope with an angle  $\gamma$ of approximately $-50^\circ$. More generally, the slope angle ($\gamma)$ is defined as follows:

\begin{equation}
\gamma=\atan\left(\frac{\partial \alpha }{ \partial (ka)}\right),
    \label{eq:}
\end{equation}
where a  value of $\gamma=-90^\circ$ corresponds to a vertical line (narrow band) and $\gamma=0^\circ$ indicates a horizontal line (wide band).

The slope can be expressed in angular frequency derivatives and using the characteristic eigenvalues,
\begin{equation}
    \frac{\partial \alpha}{\partial (ka)}= \frac{c}{a}\frac{\partial \alpha}{\partial \omega}= \frac{c\alpha'}{a}=-\frac{c}{a}\frac{\lambda'}{1+\lambda^2},
    \label{eq:alphalambda}
\end{equation}
where prime $\left ('\right )$ notation indicates a derivative with respect to the angular frequency. Alternatively, by differentiating the characteristic mode formulation in \eqref{eq:CM} with respect to angular frequency, we obtain for finite $\lambda_n$~(see App.~\ref{app:1}) 
    \begin{equation}
    \lambda_n' =\frac{\mat I_n^\mrm H (\mat X'-\lambda_n \mat R')\mat I_n}{\mat I^\mrm H_n\mat R \mat I_n},
    \label{eq:lambdaprime}
\end{equation}
where the superscript ${}^{\herm}$ denotes Hermitian transpose.

\begin{figure}
    \centering
    \includegraphics[width=0.95\linewidth]{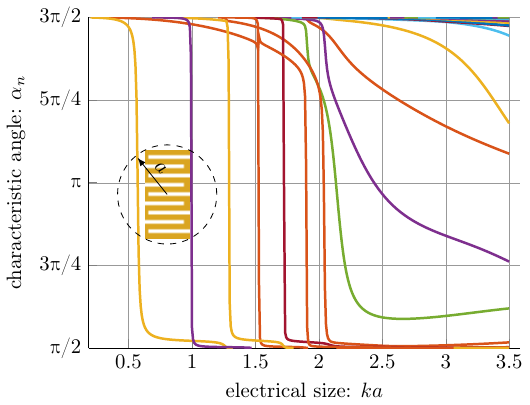}
    \caption{Characteristic angles of a PEC meander-type realization $\varOmega_\mathrm r$, fitting within a 2:1 rectangular design region $\reg$, and circumscribed by a sphere of radius $a$.}
    \label{fig:meander}
\end{figure}

The 2:1 PEC plate is just one specific realization $\reg_\T r$ within the design region $\reg$. Many other possible designs can be considered, each exhibiting distinct performance characteristics. Notably, miniaturized structures can be designed to achieve a lower resonance frequency. To illustrate this, a meander-type realization $\reg_\T r$ within the same design region $\varOmega$ is shown in Fig.~\ref{fig:meander}. The realization exhibits no initially inductive modes within the electrical sizes shown, as it effectively behaves as a single long folded line.  Several modes resonate, with four occurring below the first resonance of the 2:1 PEC plate. However, the slope of the characteristic angles is extremely steep, approaching a nearly vertical angle of approximately $-90^\circ$. This illustrates the well-known trade-off of miniaturization; while it enables a lower resonance frequency, it comes at the cost of a narrow bandwidth~\cite{Volakis+etal2010}. 

With a multitude of designs and some known rules and trade-offs of antenna design a key question arises: What can be achieved within a design region $\reg$? Are there fundamental limits to how the characteristic modes (CMs) of a realization $\reg_\T r \subset \reg$ can behave? In particular, can a mode be engineered to remain broadband, \ie maintain $\alpha=0$ across a wide range of electrical sizes (frequencies)? 

This paper shows that the slope of the CMs is fundamentally bounded, and these bounds can be determined for any design region $\varOmega$ without explicitly specifying the realizations $\reg_\T r$ within it.  

\section{Characteristic mode bounds at resonance}
\label{sec:resonance}

A mode is resonant at $\lambda=0$ (or equivalently $\alpha = \pi$), and we expect a fundamental connection between resonance, bandwidth, and the Q factor ($Q$)~\cite{Yaghjian+Best2005,Volakis+etal2010}. This is demonstrated through the angular frequency derivative of the characteristic angles (non-degenerate), \eqref{eq:alphalambda} and~\eqref{eq:lambdaprime}, where we obtain~\cite{Harrington+Mautz1972}
\begin{equation}
    \left.\frac{\partial \alpha_n}{\partial \omega}\right \rvert_{\alpha_n=\pi}=\left.-\frac{\partial \lambda_n}{\partial \omega}\right \lvert_{\lambda_n=0}=-\frac{\mat I_n^\mrm H \mat X '\mat I_n}{\mat I_n^\mathrm H \mat R \mat I _n}=-\frac{2Q_n}{\omega},
    \label{eq:alphaQ}
\end{equation}
where $Q_n$ is the modal Q-factor. Since bandwidth is inversely proportional to $Q$~\cite{Yaghjian+Best2005,Volakis+etal2010}, this well-known result establishes a fundamental link between the variations of characteristic angles and achievable fractional bandwidth $\propto Q^{-1}$. Note that the matrix $\mat X'$ in~\eqref{eq:alphaQ} is indefinite~\cite{Gustafsson+etal2012a} for electrically large structures and that the $Q_n$ is not inversely proportional to the fractional bandwidth for these cases.
    
From fundamental bounds on antenna design, it is well-known that antenna bandwidth is inherently finite~\cite{Chu1948,Gustafsson+etal2015b}. Consequently, characteristic modes cannot remain at $\alpha = \pi$ indefinitely as frequency varies. Furthermore, it is always possible to generate resonances at arbitrarily low frequencies, resulting in low bandwidth and there is no upper bound on how large $\lambda'$ can become (equivalently no lower bound for $\alpha'$). This can partly be observed in the modes displayed in Fig.~\ref{fig:meander} where the modes exhibit a very sharp downward slope, almost vertical, compared to those in Fig.~\ref{fig:2t1plateModes}. However a minimal value for $\lambda'$ exists (equivalently a maximal value of $\alpha'$). 

Regardless of whether a specific realization $\reg_\T{r}\subset\reg$ has a particular CM mode resonant at a specific frequency, we can describe the maximal value $\alpha'$ can have at that specific frequency by solving an optimization problem over an angular frequency-scaled derivative,
\begin{equation}
\begin{aligned}
	& \minimize_{\reg_\T r} && \omega\lambda_n'\\
	& \subto && \lambda_n = 0\\
    & && \lambda_n \quad \text{CM of }\reg_\T{r}\subset\reg,
\end{aligned}
\label{eq:CMopt}
\end{equation}
where the optimization is over all possible realizations $\reg_\T{r}$ fitting within $\varOmega$. To simplify notation we use $\lambda$ instead of $\alpha$ in the analysis. This optimization problem is simple in expression; for specific characteristic eigenvalue at resonance find the minimal derivative (in extension slope), yet hard to solve. Utilizing the MoM formulation~\eqref{eq:alphaQ} and normalizing the denominator $\M{I}_n^{\herm}\M{R}\M{I}_n=1$ to unity, we reformulate~\eqref{eq:CMopt} as,
\begin{equation}
\begin{aligned}
	& \minimize_{\reg_\T{r}} && \omega \lambda_n'=\omega \M{I}_n^{\herm}\M{X}'\M{I}_n\\
	& \subto && \M{I}_n^{\herm}\M{X}\M{I}_n = 0 \\
    &  && \M{I}_n^{\herm}\M{R}\M{I}_n=1\\
    & && \M{I}_n \text{ CM current of }\reg_\T{r}\subset\reg,
\end{aligned}
\label{eq:CMoptJXw}
\end{equation}
where $\M{I}_n$ denote CM currents for any realization $\reg_\T r$ fitting within the design region $\reg_\T r\subset\varOmega$. 
The problem can be relaxed to a quadratically constrained quadratic program (QCQP) formulated over the current~\cite{Gustafsson+Nordebo2013,Capek+etal2017b}. This is done by dropping the constraint of $\M{I}_n$ being a CM current
\begin{equation}
\begin{aligned}
	& \minimize_{\M{I}} && \omega \lambda_\mathrm {b}'=\omega \M{I}^{\herm}\M{X}'\M{I}\\
	& \subto && \M{I}^{\herm}\M{X}\M{I} = 0 \\
	& && \M{I}^{\herm}\M{R}\M{I} = 1,
\end{aligned}
\label{eq:CMoptJRes}
\end{equation}
where $\lambda'_n\geq \lambda_\mathrm{b}'$, subscript indicating the lower bound and $\M{I}$ represents any current in $\varOmega$. This is recognized as a QCQP for lower bounds on Q-factors, $Q_\mathrm b$, for antennas~\cite{Capek+etal2017b,Gustafsson+etal2016a} yet again highlighting a connection between antenna performance in terms of Q-factors with the slope of CM traces. 

Here, the matrices represent the design region $\reg$, and $\M I$ denotes any current within that region. This allows us to solve for a resonant mode at any frequency of interest. The optimization solution thus establishes fundamental performance constraints for any realization $\reg_\T r$ within the design region $\varOmega$, without prescribing or providing them. These bounds reveal the inherent limitations and potential of the design region.

There are several approaches to addressing this optimization problem~\cite{Park+Boyd2017}. In this work, provided $\M X' \succeq 0$, it is reformulated as a maximization of the radiated power~\cite{Gustafsson+etal2019}
\begin{equation}
\begin{aligned}
    \maximize_\M{I} \quad & \mat I^\mathrm H\mat R \mat I\\
    \subto \quad& \mat I^\mathrm H\mat X\mat I=0\\
    & \omega \mat I^\mathrm H\mat X'\mat I=1
\end{aligned}
\end{equation}
This alternative formulation emphasizes the maximization of radiated power under the constraint that the current distribution is self-resonant, while the normalization condition ensures a fixed rate of change with respect to angular frequency. To solve this problem, we employ a dual formulation given by~\cite{Gustafsson+etal2019}
\begin{equation}
    (\omega\lambda_{\mathrm{b}})^{-1}= \min_{\nu} \max \mathrm{eig}\left(\mat U(\omega \mat X'+\nu\mat X)^{-1}\mat U^\mathrm H \right),
    \label{eq:dual}
\end{equation}
where $\mat U$ is the spherical wave matrix~\cite{Gustafsson+etal2022a} with $\mat U^\mathrm H\mat U=\mat R$ and where the Lagrange parameter $\nu$ is constrained by $\omega \mat X'+\nu \mat X \succeq \mat 0$ with range provided by the solution to $\mat X \mat I_n=\nu_n\omega \mat X'\mat I_n$ giving $\nu \in [-1/\max \nu_n, -1/\min \nu_n]$. Note that the eigenvalue problem also diagonalize the matrix ($\omega \mat X'+\nu \mat X$) which can be used to remove the matrix inversion in~\eqref{eq:dual}. 
This is a well-studied problem for antenna bounds~\cite{Capek+etal2017b,Gustafsson+etal2016a,Jelinek+Capek2017}. Cases with an indefinite $\omega\M{X}'$ are transformed to an equivalent QCQP with a positive definite object functional by using the constraint $\M{I}^{\herm}\M{R}\M{I}=1$ to add a constant $\mu=\mu\M{I}^{\herm}\M{R}\M{I}$ to the object functional in~\eqref{eq:CMoptJXw} with $\mu$ determined such that $\omega\M{X}'+\mu\M{R}\succ \M{0}$, see App.~\ref{app:2} for details.

\begin{figure}
    \centering
    \includegraphics[width=0.97\linewidth]{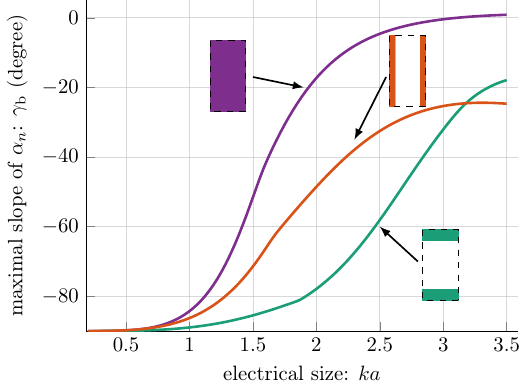}
    \caption{Maximal slope angle, $\gamma_\mathrm b$ at resonance for three design regions $\reg$ fitting within a 2:1 rectangle represented by the dashed boundary. The orange (vertical strips) and green (horizontal strips) utilize 1/3 of the total available area.}
    \label{fig:slopeResonance}
\end{figure}

By solving the optimization problem~\eqref{eq:CMoptJRes} we obtain the lower bound $\lambda'_\mathrm{lb}$ and maximal bounds  $\alpha'_\mathrm b$ and $\gamma_\mathrm b$. No characteristic mode can have a slope exceeding this limit. In Fig.~\ref{fig:slopeResonance}, this bound is shown at resonance through the maximal value of $\gamma$ for various design regions $\reg$, all within a 2:1 rectangle. For small $ka$ values (less than 1), the bound approaches $-90^\circ$, indicating very narrow bandwidth performance. As the structure size increases, the maximum value shifts toward $90^\circ$, essentially removing any significant limits (although here the limit within the presented $ka$ range is $\approx0^\circ$). It is evident that a larger structure provide more relaxed bounds compared to smaller ones, as the corresponding curves for smaller structures always fall below those of the larger ones. This is expected, as larger structures offer greater design flexibility. 

The meander line in Fig.~\ref{fig:meander} represents a design that fits within the purple design region, and the narrow-band performance of the first resonant modes is evident, as the slope must be very steep, with $\gamma < -85^\circ$ for $ka < 1$. For the particular designs considered, we observe that a single mode cannot exhibit a broadband resonance before $ka\approx2.5$ as this is where the slope angle $|\gamma_\mathrm b|<10^\circ$, with smaller regions experiencing this regime for even larger $ka$ values. Moreover, since the maximal slope is almost always negative (except for the full region at $ka>3$), any mode appearing in this range must come from the capacitive side, while initially inductive modes can only emerge and become resonant for $ka>3$.

\section{General characteristic mode bounds}
\label{sec:general}
We have so far focused only on resonance due to its fundamental connection to a well-understood bounded quantity, the Q-factor and bandwidth~\cite{Volakis+etal2010,Chu1948,Gustafsson+etal2015b}. However, there is no strict reason to impose this requirement. The parameters may still be bounded, albeit with a less direct physical interpretation. To extend the formulation, we now consider any value of $\lambda$. In general, for non-degenerate eigenvalues, consider the derivative~\eqref{eq:lambdaprime}.
It is mathematically simple to modify the optimization problems~\eqref{eq:CMopt} to~\eqref{eq:CMoptJRes} in Sec.~\ref{sec:resonance} to,
\begin{equation}
\begin{aligned}
    \minimize \quad & \omega \lambda'=\omega \mat I^\mathrm H(\mat X'-\lambda \mat R')\mat I\\
    \subto \quad& \mat I^\mathrm H(\mat X-\lambda \mat R)\mat I=0\\
    & \mat I^\mathrm H\mat R \mat I=1
\end{aligned}
\label{eq:CMoptJ}
\end{equation}
This QCQP has the same structure as the self-resonant formulation~\eqref{eq:CMoptJRes} but with different matrices. It is also solved in the same way, \eg for PD cases $\M X'-\lambda \M R'\succeq 0$
\begin{equation}
\begin{aligned}
    \maximize \quad & \mat I^\mathrm H\mat R \mat I\\
    \subto \quad& \mat I^\mathrm H(\mat X-\lambda \mat R)\mat I=0\\
    & \omega \mat I^\mathrm H(\mat X'-\lambda \mat R')\mat I=1
\end{aligned}
\end{equation}
The dual problem for the lower bound of $\lambda'$ can be formulated as
\begin{multline}
    (\omega\lambda_{\mathrm{b}}')^{-1}= \min_{\nu} \max \mathrm{eig}\\ \left (\mat U(\omega(\mat X'-\lambda_n \mat R')+\nu(\mat X-\lambda_n \mat R))^{-1}\mat U^\mathrm H \right ),
    \label{eq:dual2}
\end{multline}
where, as before, $\nu$ is provided from $\omega(\mat X'-\lambda_n \mat R')+\nu(\mat X-\lambda_n \mat R) \succeq \mat 0$ with range provided by an eigenvalue problem as in~\eqref{eq:dual}. Indefinite cases are solved similar as for the self-resonant case~\eqref{eq:CMoptJRes} by adding $\mu\M{I}^{\herm}\M{R}\M{I}$ to the object functional, see App.~\ref{app:2} for details.

This approach now allows us to determine the maximal slope of $\alpha$ for any characteristic angle at any electrical size, $ka$ (or frequency). We can graphically illustrate this fundamental bound using arrows to represent the slopes of $\alpha$, which are constrained in the \textcolor{red}{upward} direction but not in the \textcolor{dgreen}{downward} direction, \ie
\begin{center}
\begin{tikzpicture}
\draw[->] (0,0) -- (0.707,-0.707);
\draw[->,dgreen] (0,0) -- (0.6,-0.8);
\draw[->,red] (0,0) --  (0.8,-0.6);	
\node[anchor=west] at (0.4,-0.2){$\gamma_{\T{b}}$};
\node[anchor=east,dgreen] at (0,-0.3) {possible};
\node[anchor=west,red] at (0.7,-0.6) {impossible};
\draw[dashed] (0,0) -- (0.707,0);
 \draw[thick] (0.5,0) arc[start angle=0, end angle=-45, radius=0.5];
\end{tikzpicture}
\end{center}

\begin{figure}
    \centering
    \includegraphics[width=0.95\linewidth]{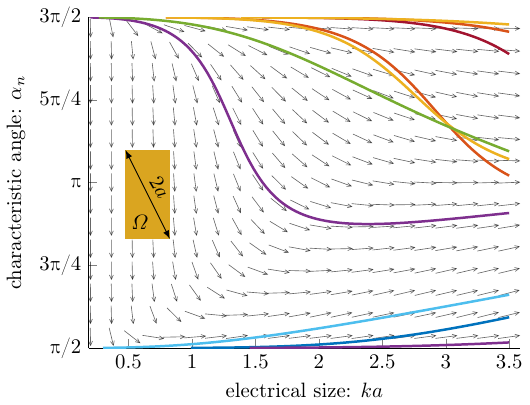}
    \caption{Characteristic angles shown in Fig.~\ref{fig:2t1plateModes} with the bounds on maximal slope of $\alpha$ from solving~\eqref{eq:CMoptJ}, displayed as arrows.}
    \label{fig:FullBounds}
\end{figure}

For the previously discussed rectangular plate, these bounds lead to the constraints shown in Fig.~\ref{fig:FullBounds}, where the characteristic modes from Fig.~\ref{fig:2t1plateModes} are also displayed.
We observe that the characteristic modes adhere to the fundamental limit of the design region $\varOmega$. The first resonant mode (purple) follows these limits tightly up until $ka\approx 2$, after which some deviation occurs. A fundamental property of antenna performance is polarizability~\cite{Gustafsson+etal2015b}, and currents that effectively allow for charge separation are situated closer to the bounds. For the PEC plate, the solid purple line, representing the dipole mode with current along the long edge at resonance, use the structure effectively, while other modes are less constrained relative to the feasible bound. However, a small deviation between realized CM values and the bounds can be found as the bounds involve a combination of electric and magnetic dipole components, which reduce the Q-factor~\cite{Chu1948,Capek+Jelinek2016,Capek+etal2017b} and consequently increase the slope of $\alpha_\mathrm b$.

\section{Illustrative examples}
\label{sec:results}
\begin{figure}
    \centering
    \includegraphics[width=0.95\linewidth]{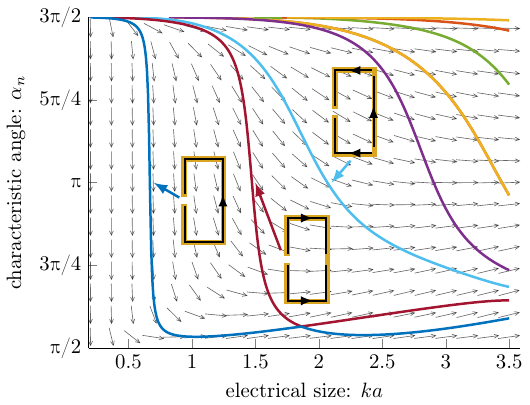}
    \caption{Characteristic angles for the modes of a split-ring realization $\reg_\T r$ circumscribed by a sphere of radius $a$. The bounds on the maximal slope of $\alpha_n$ are obtained from solving~\eqref{eq:CMoptJ} for the design region $\reg$, displayed as arrows. The inset images illustrate the path of the characteristic currents at $\alpha_n=\pi$ for the first three modes.}
    \label{fig:SplitringCM}
\end{figure}

The bound on the maximal slope of $\alpha$ applies to any realization $\reg_\T r$ fitting within the design region $\reg$ to which the bounds are valid. To illustrate this, we compute the bounds $\alpha'_\mathrm{b}$ of a 2:1 rectangular design region $\reg$ and examine the characteristic modes from various realizations $\reg_\T r\subset\reg$. In Fig.~\ref{fig:SplitringCM}, a simple split ring geometry and its associated characteristic modes are shown. The bounds are tightest for the first mode, which is the dipole mode, as indicated by the inset image, where a schematic illustration of the current path is overlaid. The structure produces optimal charge separation, aligning with the utilization of the structure's polarizability. The other modes are less tightly constrained, but they never exceed the limits. Similar to the meander structure in Fig.~\ref{fig:meander}, this folded design removes any onset of inductive modes within the considered electrical size range.

\begin{figure}
    \centering
    \includegraphics[width=0.95\linewidth]{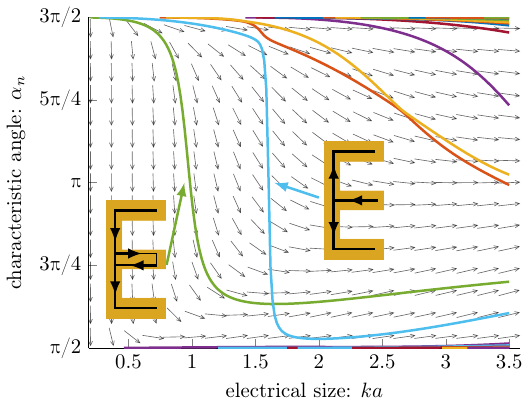}
    \caption{Characteristic angles for the modes of an E-shape realization $\reg_\T r$. The bounds on the maximal slope of $\alpha_n$ are obtained from solving~\eqref{eq:CMoptJ} for the design region $\reg$, displayed as arrows. The inset images illustrate the path of the characteristic currents at $\alpha_n=\pi$ for the first two modes.}
    \label{fig:EshapeCM}
\end{figure}

In Fig.~\ref{fig:EshapeCM}, a region shaped like the letter ``E'' is considered. The first resonant mode, occurring around $ka = 1$, takes advantage of the altered boundary to achieve a resonance frequency lower than that of the 2:1 plate. However, this comes at the cost of a reduced bandwidth. Nevertheless it still aligns well with the bounds for the larger region. The second mode, on the other hand, shows current distributions that are less favorable in terms of polarizability, leading to very narrow-band performance that is considerably sharper than the fundamental limit set by the region. Despite this, the higher-order modes exhibit a better fit, showing less restrictive performance and more favorable characteristics in relation to the bounds.

\begin{figure}
    \centering
    \includegraphics[width=0.95\linewidth]{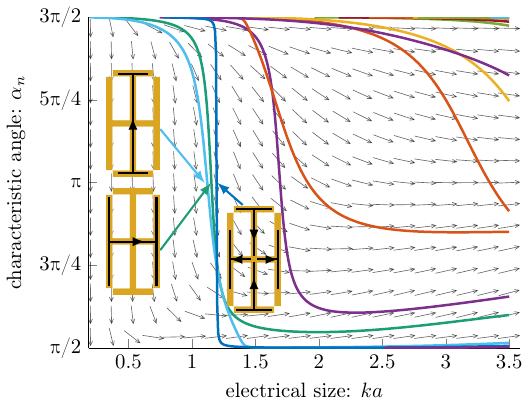} 
    \caption{Characteristic angles for three modes for a crosspotent-shape realization $\reg_\T r$. The bounds on the maximal slope of $\alpha_n$ are obtained from solving~\eqref{eq:CMoptJ} for the design region $\reg$, displayed as arrows. The inset images illustrate the path of the characteristic currents at $\alpha_n=\pi$ for the first three modes.}
    \label{fig:CrosspotentCM}
\end{figure}

Another demonstrative example is the cross-potent design shown in Fig.~\ref{fig:CrosspotentCM}. The first two resonances correspond to similar current distributions that follow the limits represented by the arrows quite well. However, the third mode utilizes the structure's polarizability much less effectively, resulting in a very narrow-band performance with a significantly steeper slope than the limit, producing the intriguing outcome of three resonances occurring near the same $ka$. In this design, initial inductive modes is not possible within the range.

Building on these results, we find that it is impossible, by our definition, to produce an initial inductive mode reaching resonance within this region, as the fundamental limit does not permit it in the considered range. Consequently, all resonant modes of the miniaturized designs must arise from the capacitive region. The bounds indicate that for small electrical sizes, resonances within this region exhibit a large negative slope, characteristic of capacitive modes. As expected, no single mode demonstrates wideband performance, and inductive modes are not observed, as achieving them would require a region where $\gamma_\mathrm b> 0$

The results further demonstrate that all designs adhere to the established constraint. While the bound may be less tight when only a small portion of the available region is utilized, it remains valid in all cases.

\begin{figure}
    \centering
    \includegraphics[width=0.97\linewidth]{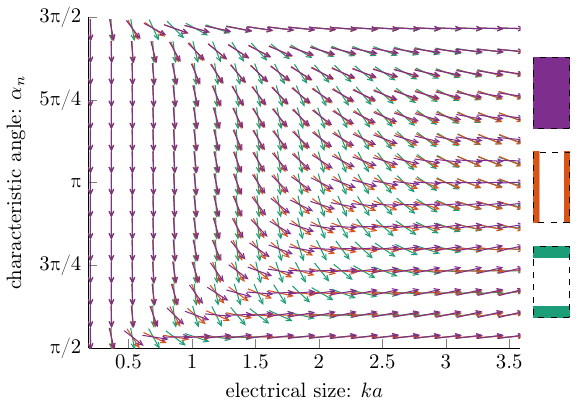}
    \caption{Bounds on the slope, $\alpha_\mathrm b'$, presented as arrows for three regions within a 2:1 plate (highlighted by a dashed contour) and circumscribed by a sphere of radius $a$. Both the orange and green regions occupy $1/3$ of the area of the purple region.}
    \label{fig:ManyBounds}
\end{figure}

To illustrate the impact of altering the available region, Fig.~\ref{fig:ManyBounds} presents the bounds for the regions in Fig.~\ref{fig:slopeResonance}. For $ka < 1$, the effect is visually minimal due to the steep slope in this range. The primary impact of reducing the region is observed above the first resonance of the bounding structure, occurring at $ka = \sqrt{5}\pi/4 \approx 1.76$, though the shift begins slightly before this point. Additionally, as the region shrinks, the resulting bound becomes increasingly restrictive, as evidenced by the downward trend of the arrows. Both the limits of the orange design (vertical strips) and green design (horizontal strips) are bounded by the purple structure (full).

\section{Higher order modes}\label{Sec:HOM}
The presented bounds on the characteristic mode (CM) slopes are valid for all structures within the considered design region. The lowest-order (first) modes closely follow these bounds, as seen in Fig.~\ref{fig:FullBounds} to~\ref{fig:CrosspotentCM}. However, higher-order modes tend to exhibit steeper slopes, as illustrated by the intersecting modes in Fig.~\ref{fig:CrosspotentCM}. Tighter bounds for these higher-order modes can be derived similarly to~\eqref{eq:CMoptJRes} and~\eqref{eq:CMoptJ}, by leveraging the orthogonality~\cite{Harrington+Mautz1971} of the CMs, \ie $\M{I}_n^{\herm}\M{R}\M{I}_m = \delta_{nm}$ and $\M{I}_n^{\herm}\M{X}\M{I}_m = \lambda_m \delta_{nm}$, where $\delta_{nm}$ denoted the Kronecker delta. Bounds on higher-order modes provide insight into the number of simultaneous usable modes within a given design region, \eg useful for MIMO~\cite{Li+etal2014,Yang+Adams2015,Peitzmeier+Manteuffel2019} and similar to the concept of degrees of freedom for CM~\cite{Gustafsson+Lundgren2024}.

By determining the bounds iteratively and enforcing orthogonality, we obtain improved bounds for higher-order modes, particularly in cases where the lower-order modes already lie close to their bounds at the given frequency. The relaxed optimization problems~\eqref{eq:CMoptJRes} and~\eqref{eq:CMoptJ} do not enforce that the optimal current is a CM, and hence do not guarantee simultaneous orthogonality with respect to both $\M{R}$ and $\M{X}$. To address this, we enforce orthogonality with respect to $\M{R}$ iteratively, by introducing the affine constraint $\M{I}_{\T{b}n}^{\herm}\M{R}\M{I}_{\T{b}m} = 0$ for $n > m$ into~\eqref{eq:CMoptJRes} and~\eqref{eq:CMoptJ}, where $\M{I}_{\T{b}m}$ denotes the optimal current at iteration $m$.

This constraint is implemented by reducing the matrix sizes in~\eqref{eq:CMoptJRes} and~\eqref{eq:CMoptJ}, projecting them onto the subspace $\M{W}$ orthogonal to $\M{R}\M{I}_{\T{b}m}$, \eg via $\M{R} \to \M{W}^{\herm}\M{R}\M{W}$. The resulting optimization problems are structurally identical to~\eqref{eq:CMoptJRes} and~\eqref{eq:CMoptJ}, but involve reduced matrices and are solved in the same manner. Finally, it is crucial to consider degenerate eigenvalues, especially for symmetric geometries~\cite{Capek+etal2017b}, to ensure the absence of a duality gap and to guarantee that the current $\M{I}_{\T{b}n}$ is a valid solution to the optimization.

\begin{figure}
    \centering
    \includegraphics[width=1\linewidth]{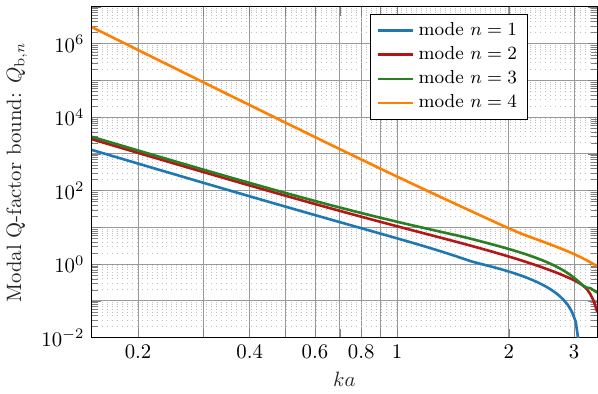}
    \caption{Lower bound~\eqref{eq:alphaQ} on the modal Q-factors $Q_{\T{b}n}$ based on orthogonal far fields for the four lowest CM at resonance for a $2:1$ rectangular design region.}
    \label{fig:Qlb}
\end{figure}

The resulting bounds on the four lowest modal Q-factors, $Q_{\T{b}n}$ for $n = 1, 2, 3, 4$,  for a 2:1 rectangular design region $\reg$ are shown in Fig.~\ref{fig:Qlb}. The lowest bound coincides with the fundamental lower bound on the Q-factor established in~\cite{Capek+etal2017b, Gustafsson+etal2019}, and it scales as $(ka)^{-3}$ for $ka \ll 1$. This mode is primarily dominated by an electric dipole oriented along the length of the structure for $ka \ll 1$. The second mode exhibits a higher Q-factor and corresponds to an electric dipole along the shorter dimension of the plate. The third mode is primarily characterized by a loop current. However, all these modes involve a mixture of electric and magnetic dipole components, which can collectively reduce the Q-factor~\cite{Chu1948}. The fourth mode corresponds to a quadrupole, and its Q-factor scales as $(ka)^{-5}$~\cite{Collin+Rothschild1964}. The modal Q-factors are only plotted for positive values, and the inverse proportionally of $Q_n$ to fractional bandwidth is only valid for $Q_n\gg 1$.

\begin{figure}
    \centering
       \includegraphics[width=1\linewidth]{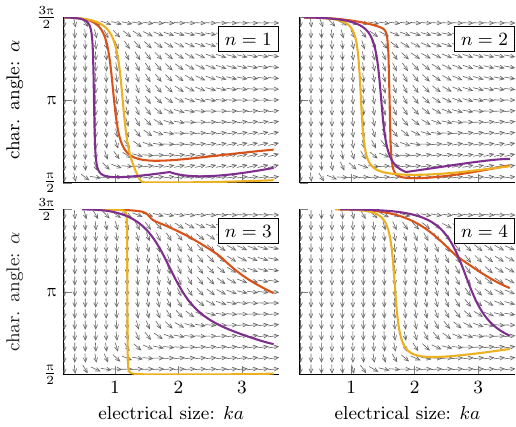}
    \caption{Bounds on the slope of the first four characteristic angles $\alpha_{\T{b}n}$ for a $2:1$ rectangular design region $\reg$. Characteristic angles $\alpha_n$ for $n=1,2,3,4$ of the split ring (purple), E-shape (orange), and crosspotent (yellow) are included for comparison.}
    \label{fig:QlbArrow}
\end{figure}

Bounds on the slope of the characteristic angle for the first four modes in a rectangular design region are shown in Fig.~\ref{fig:QlbArrow}. The resonant cases, where $\alpha_n = \pi$, correspond to the data presented in Fig.~\ref{fig:Qlb}. In particular, the high Q-factor of the fourth mode, $Q_{\T{b}4}$, is associated with a steep slope in the characteristic angle.

Characteristic angles $\alpha_n$ of the split ring (purple, Fig.~\ref{fig:SplitringCM}), E-shape (orange, Fig.~\ref{fig:EshapeCM}), and crosspotent (yellow, Fig.~\ref{fig:CrosspotentCM}) are superimposed in the plots for comparison. Note that the bounds (indicated by arrows) are derived iteratively for orthogonal modes at the same frequency. As a result, higher-order bounds can be used as guidelines for the higher-order characteristic modes (CM), but they do not strictly constrain their slopes. 

\section{Conclusion}
\label{sec:conc}
In conclusion, this work has established fundamental bounds on the slope (frequency derivative) of characteristic mode eigenvalues, demonstrating their universal applicability within a given design region. The derived bounds can effectively describe the behavior of the first modes up to a certain electrical size, particularly for small $ka$, where they provide tight constraints on modal evolution. These fundamental limits are formulated as a quadratically constrained quadratic program (QCQP) and are solved through the dual problem and apply to all realizations within the design region, enabling the analysis of potential structures within a specified region and background environment.

Through examples such as the split ring, E-shape, and crosspotent designs we have illustrated the practical implications of these bounds. In these cases, the first resonant modes exhibit behavior consistent with the derived constraints, with the E-shaped region showing narrow-band performance at higher modes. These examples highlight how the bounds apply across different design approaches, providing valuable insight into their real-world relevance.

The first resonant modes adhere closely to these constraints. In contrast, higher-order modes tend to exhibit steeper slopes, prompting the need for tighter bounds. To address this, the framework was extended by iteratively enforcing orthogonality through the inclusion of an affine constraint in the optimization, which leads to improved bounds for higher-order modes. This iterative orthogonalization refines the limits on modal slopes but also provides insight into the number of simultaneously usable modes, a detail with important implications for MIMO applications and the analysis of modal degrees of freedom. It is important to emphasize that these higher-order mode bounds presented here should be interpreted as practical guidelines rather than strict constraints. The bounds are valid under the assumption that all preceding lower-order modes have already reached their respective bounds, which are computed for each frequency and specific $\alpha$ value. Despite some deviation from these assumptions in the presented examples, the result nonetheless align well with the corresponding higher-order mode bounds. 

The derived bounds act as a feasibility criterion, indicating whether a given mode can theoretically achieve a specific performance. However, they do not confirm whether a realizable antenna design can attain these modal characteristics. Moreover, the results are specific to individual characteristic modes and do not account for potential modal interactions or combinations in practical designs. Nevertheless, antennas remain subject to similar fundamental limitations~\cite{Gustafsson+etal2007a, Gustafsson+Nordebo2013, Capek+etal2017b, Gustafsson+Lundgren2024}. The bounds can be visualized through arrow plots, which represent flow lines along which the characteristic angles cannot increase at a faster rate. This provides insights into achievable bandwidth for antenna structures and explicitly illustrates the cost of miniaturization.

While the presented results and examples focus on surface PEC objects, the method for deriving these bounds extends to three-dimensional structures composed of any lossless material, subject to the specifics of the MoM implementation. This work can be extended to a substructure environment, which will be explored in future research.

\appendices
\section{Angular frequency derivative of the characteristic eigenvalues}
\label{app:1}
Starting from the generalized eigenvalue equation~\eqref{eq:CM} we take the angular frequency derivative~\cite{Petersen+Pedersen2012},
\begin{equation}
    \frac{\partial}{\partial \omega}\left ( \mat X \mat I_n \right )=\frac{\partial}{\partial \omega}\left ( \lambda_n \mat R \mat I_n \right ).
    \end{equation}
    Applying the product rule gives
    \begin{equation}
    \mat X' \mat I_n +\mat X \mat I'_n = \lambda_n' \mat R \mat I_n+ \lambda_n \mat R' \mat I_n +\lambda_n \mat R \mat I'_n ,
    \end{equation}
    with prime ($'$) noting the angular frequency derivative. Multiply both sides on the left with $\mat I^\mrm H$, we obtain
    \begin{equation}
    \mat I_n^\mrm H \mat X' \mat I_n +\mat I_n^\mrm H\mat X \mat I'_n = \lambda_n' \mat I_n^\mrm H \mat R \mat I_n+ \lambda_n \mat I_n^\mrm H \mat R' \mat I_n +\lambda_n \mat I_n^\mrm H \mat R \mat I'_n.
    \label{eq:lambdaderiv}
    \end{equation}
    
    Using a rearranged form of the eigenvalue equation \eqref{eq:CM},
    \begin{equation}
   \mat I_n^\mrm H \mat X = \lambda_n \mat I_n^\mrm H \mat R,
    \end{equation}
    we can eliminate (subtract) the two terms containing $\mat I_n'$ from~\eqref{eq:lambdaderiv} and solve for $\lambda_n'$,
    \begin{equation}
    \lambda_n' =\frac{\mat I_n^\mrm H (\mat X'-\lambda_n \mat R')\mat I_n}{\mat I^\mrm H_n\mat R \mat I_n},
    \end{equation}
    with the assumption that the structure is radiating, $\mat I_n^\mrm H \mat R \mat I_n> 0$.
    
\section{QCQP}
\label{app:2}
The QCQPs~\eqref{eq:CMoptJRes} and~\eqref{eq:CMoptJ} are solved using the parametrized eigenvalue problems~\eqref{eq:} in cases where the objective functionals are positive definite (PD), such as when $\omega\M{X'}\succ\M{0}$ in~\eqref{eq:CMoptJRes}. This reformulation relies on transforming the minimization of $\M{X}_{\T{w}}=\omega\M{X}'$ into a maximization of the radiated power, which necessitates a PD functional.

For indefinite cases, we add a constant via the constraint $\M{I}^{\herm}\M{R}\M{I}=1$ into the objective functional, thereby constructing a new QCQP with a PD objective functional. Notably, adding this constant alters only the function’s value, not the resulting optimal current $\M{I}$.

To ensure a PD objective functional, we introduce a constant $\mu\geq 0$ such that $\M{X}_{\T{w}}+\mu\M{R}\succ \M{0}$. This condition can be expressed as $\mu\geq -\min\eig(\M{X}_{\T{w}},\M{R})=-\min{\nu_n}$, where $\nu_n$ corresponds to the generalized eigenvalue problem $\M{X}_{\T{w}}\M{I}_n=\nu_n\M{R}\M{I}_n$. Further simplification leads to a standard eigenvalue problem by rewriting it as $\nu_n^{-1}\M{I}_n=\M{X}_{\T{w}}^{-1}\M{R}\M{I}_n$. Using the decomposition $\M{R}=\M{U}^{\herm}\M{U}$, we obtain $\M{U}\M{X}_{\T{w}}^{-1}\M{U}^\T{H}\M{f}_n=\nu_n^{-1}\M{f}_n$.


\begin{IEEEbiography}[{\includegraphics[width=1in,height=1.25in,clip,keepaspectratio]{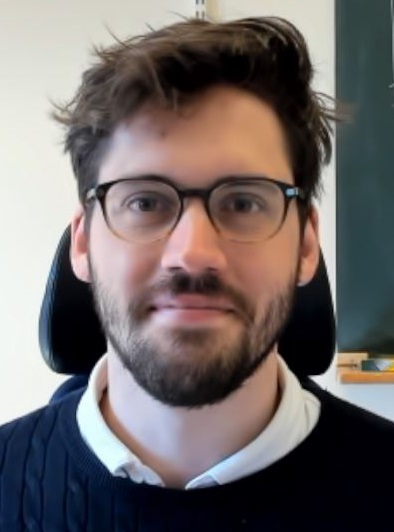}}]{Johan Lundgren} (M'22) is an assistant professor at Lund University. he received his M.Sc degree in engineering physics 2016 and Ph.D. degree in Electromagnetic Theory in 2021 all from Lund University, Sweden. 

His research interests are in electromagnetic scattering, wave propagation, computational electromagnetics, characteristic modes, functional structures, meta-materials, inverse scattering problems, imaging, and measurement techniques.
\end{IEEEbiography}

\begin{IEEEbiography}[{\includegraphics[width=1in,height=1.25in,clip,keepaspectratio]{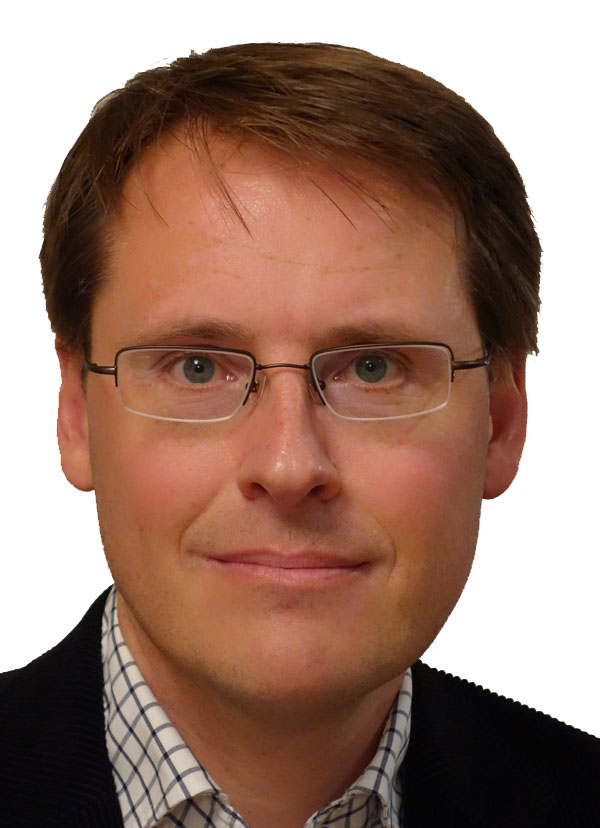}}]{Mats Gustafsson}
received the M.Sc. degree in Engineering Physics 1994, the Ph.D. degree in Electromagnetic Theory 2000, was appointed Docent 2005, and Professor of Electromagnetic Theory 2011, all from Lund University, Sweden.

He co-founded the company Phase Holographic Imaging AB in 2004. His research interests are in scattering and antenna theory and inverse scattering and imaging. He has written over 100 peer-reviewed journal papers and over 100 conference papers. Prof. Gustafsson received the IEEE Schelkunoff Transactions Prize Paper Award 2010, the IEEE Uslenghi Letters Prize Paper Award 2019, and best paper awards at EuCAP 2007 and 2013. He served as an IEEE AP-S Distinguished Lecturer for 2013-15.
\end{IEEEbiography}

\end{document}